# Enhancement of Ion Diffusion by Targeted Phonon Excitation

*Kiarash Gordiz[1], Sokseiha Muy[2], Wolfgang G. Zeier[3], Yang Shao-Horn[1,2,4], and Asegun Henry[1]\*\**

[1]Department of Mechanical Engineering, Massachusetts Institute of Technology, Cambridge MA, 02139, USA

[2]Department of Materials Science and Engineering, Massachusetts Institute of Technology, Cambridge MA, 02139, USA

[3]Institute for Inorganic and Analytical Chemistry, University of Muenster, Correnstr. 30, 48149 Münster, Germany

[4]Research Laboratory of Electronics, Massachusetts Institute of Technology, Cambridge MA, 02139, USA

\*\***Correspondence to:** ase@mit.edu

# Abstract

Ion diffusion is important in a variety of applications, yet fundamental understanding of the diffusive process in solids is still missing, especially considering the interaction of lattice vibrations (phonons) and the mobile species. In this work, we introduce two formalisms that determine the individual contributions of normal modes of vibration (phonons) to the diffusion of ions through a solid, based on (*i*) Nudged Elastic Band (NEB) calculations and (*ii*) molecular dynamics (MD) simulations. The results for a model ion conductor of Ge-substituted $Li_3PO_4$ ($Li_{3.042}Ge_{0.042}P_{0.958}O_4$) revealed that more than 87% of the $Li^+$ ion diffusion in the lattice originated from a subset of less than 10% of the vibrational modes with frequencies between 8 and 20 THz. By deliberately exciting a small targeted subset of these contributing modes (less than 1%) to a higher temperature and still keeping the lattice at low temperature, we observed an increase in diffusivity by several orders of magnitude, consistent with what would be observed if the entire material (i.e., all modes) were excited to the same high temperature. This observation suggests that an entire material need not be heated to elevated temperatures to increase diffusivity, but instead only the modes that contribute to diffusion, or more generally a reaction/transition pathway, need to be excited to elevated temperatures. This new understanding identifies new avenues for increasing diffusivity by engineering the vibrations in a material, and/or increasing diffusivity by external stimuli/excitation of phonons (e.g., via photons or other interactions) without necessarily changing the compound chemistry.

# Introduction

Solid state mass and ion diffusion is central to many applications ranging from batteries[1] and fuel cells[2] to sensors[3] and filters.[4] In many of these applications, the performance is limited by the slow diffusion of mass or ions. While some families of solid ion conductors, such as silver iodide and $Li_{10}GeP_2S_{12}$ exhibit high ion conductivities of ~1 S/cm[5] and ~0.01 S/cm [6, 7] at room temperature respectively, achieving high ion conductivities for divalent ions including oxides is challenging. For the conductivity of oxygen ions exhibited by yttria stabilized zirconia[8, 9] to reach 0.01 S/cm, an elevated temperature of 1100 K is needed. Higher oxygen diffusivity at a lower operating temperature could enable a lower system cost, longer life and greater proliferation.[10] Great advances have been made to increase the ion transport by cation and anion substitution to alter the charge carrier density and the diffusion pathways to facilitate low energy jumps. The classical approaches of chemical modification such as cation and anion substitution have led to tunable conductivity over several orders of magnitude in several families of super-ionic conductors, such as LISICON[6, 11] and NASICON.[12, 13] Thus, regardless of the mobile ion species, understanding the key factors that dictate ion conductivity is needed to devise new strategies to further increase the conductivity beyond tuning of chemical compositions.

One possible avenue to improve diffusivity without modifying the chemistry stems from considering the local ion movement in the structure. At a given temperature, ions thermally vibrate in their specific crystallographic lattice site until sufficient thermal energy is available for an ion jump. The coupled ion vibration and ion movement that is paramount for ion diffusion leads to the idea that lattice vibrations (phonons) may play a role in the ion jump. Such an idea can further be intuited by considering the role of temperature as a spectrum of scalar contributions, not as a single scalar value. Based on this more detailed view, in solids and rigid molecules, temperature is composed of a summation of individual contributions by modes of vibration. By mode of vibration, it is meant that a group of vibrating atoms can be described as a summation of collective vibrations, each with a specific frequency, often termed eigen modes, normal modes or phonons. According to the traditional description for solid state diffusion based on the Arrhenius relation $D \propto \exp(-E_a/k_B T)$,[14] diffusivity increases exponentially with increasing temperature, where $k_B$ and $T$ denote the Boltzmann's constant and the bulk temperature. The term "bulk" temperature here denotes the temperature that would be sensed if the material were in thermal equilibrium. While temperature is generally understood as a proxy for the level of excitation of the phonons or normal modes of vibration within a solid material, it should be noted that each mode has its own time varying amplitude and individual temperature. It is possible for a material to experience some non-equilibrium in the individual mode temperatures, whereby certain modes are highly excited to an effective temperature above all others (e.g., joule heating of optical modes in a transistor[15, 16]). In such a situation, a subset of modes can behave as though they are at a higher effective temperature than the bulk temperature.

In the classical limit, for an individual mode labeled $n$, the modal temperature $T_n$ is obtained from $k_B T_n = Q_n^2 \omega_n^2$, where $Q_n$ and $\omega_n$ are the modal displacement coordinate[17] (which can be interpreted as the mode amplitude) and frequency of vibration of mode $n$ (see the ESI for a derivation of this relation and its quantum analog). The idea of modal temperatures have also been introduced in previous theoretical studies,[18] where for the first time the two-temperature model was extended to study the weak coupling between certain groups of phonons and its effect on thermal transport in solid materials. Our focus in this study is the distinction between the bulk and modal temperatures, and how it influences solid-state ion transport. As the "bulk" temperature increases, the amplitudes of all the modes in the system increase, which facilitates ions to hop over activation barriers, thereby enhancing diffusion. There are practical ways, however, to increase the amplitude and temperature of a small subset of normal modes by exciting them with light either directly through photon-phonon coupling[19-22] or indirectly through electron-photon coupling[23-25] or THz electric fields[26] and keeping them in non-equilibrium with respect to the rest of the modes, without increasing the "bulk" temperature. This idea of selective vibrational mode excitation has been utilized in previous studies, as a novel tool, to modify the electronic properties of the structure through electron-phonon coupling for plasmonics,[27] photovoltaics[28, 29] and charge transport applications.[30] Selective modal excitations have also been utilized to induce vibrational and structural changes in the system for phase-change applications.[31, 32] Inspired by recent experimental observation of solid-state ion diffusion induced by external THz-electric field illumination,[26] in this study, we aim to examine if only a small subset of modes in the lattice are responsible for the hopping of ions, which would allow the increase of the diffusivity by exciting these subset of modes without increasing the bulk temperature. Experimental realization of such an idea would have far-reaching implications for chemical reactions, phase transitions or other related phenomena, because it would suggest that migration along a reaction/transition coordinate could be accelerated by only exciting the most important normal modes, rather than all modes via an increase in the bulk temperature. This approach would mean that a material and its container could remain "cold" but it could exhibit kinetics associated with a much higher temperature, by only making a subset of the most important modes "hot". To assess this opportunity, we use the example of ion diffusion. We (*i*) identify the mode level contributions to ion diffusion to determine how many modes are responsible for an ion hop in the system, and (*ii*) quantify to what extent diffusivity can be increased by targeted excitation of a small subset of these contributing modes without increasing the bulk temperature.

It is important to recognize that the role of specific structural modes and phonons in facilitating solid-state diffusion has been reported in other studies.[33-44] The frequency of longitudinal acoustic phonons was shown to exhibit a strong correlation with the enthalpy of self-diffusion in body-centered cubic metals.[33] In addition, octahedral rotation has been shown to promote fast oxygen ion conduction in perovskite-related phases.[34-38] Li and Benedek[39] calculated the force constants associated with octahedral rotations in different Ruddlesden−Popper phases and showed that the willingness of a material to go under these octahedral rotations correlates with the migration barrier of oxygen ions. Similarly, anion rotation in closo-Borate solid electrolytes has been shown to be important for ion transfer through the lattice.[40-42] More recently, Muy et al[43] have shown that lowering the computed average vibrational frequency of lithium sublattice in LISICON by cation and anion substitution is accompanied with reduced migration barrier. Similar correlations between activation energy and computed average frequency of sodium ions have been noted for sodium ion conductors,[44] where activation energy is correlated with the Debye temperature, a proxy for the softness of the lattice, i.e. corresponding to low energy vibrational modes. However, the average frequency of vibration reflects a weighted sum over all phonon modes, and the Debye temperature only measures the slopes of the acoustic phonon modes at very low wave vectors, neglecting the influence of the optical phonons and including the effect of some irrelevant phonons to ion migration. Most of these insights have been obtained from density-functional theory (DFT) lattice relaxation, *ab-initio* molecular dynamics (AIMD), or experimental characterization. However, no direct information about specific phonons and their respective eigenvectors have been employed in a systematic way to obtain a deeper insight into these phenomena. The information for eigenvectors of phonons have been used in several modal analysis methodologies for thermal transport including atomistic Green's function (AGF),[45, 46] Boltzmann transport equation (BTE)[47, 48] and molecular dynamics (MD) simulations.[49-52] These methods have shown that in different materials, certain group of modes could contribute to heat transfer with higher rates, for instance, the high frequency in-plane phonons in graphene[53] and interfacial modes in semiconductor interfaces.[54-56] In this study, we aim to identify modes of vibration that facilitate the ion migration by directly using the knowledge of phonons and their respective eigenvectors, based on the approaches that were recently developed in the context of MD-based phonon analysis.[49-52]

Here, we chose Ge-substituted $Li_3PO_4$ as our model system, since $Li_3PO_4$ is a well-known parent structure, from which numerous fast solid conductors have been developed including LGPS,[57, 58] and its ion conductivity can be tuned up to ~10 orders of magnitude by aliovalent and anion substitution.[7] To identify the modal contributions to lithium ion diffusion we introduce two methodologies, one based on nudged elastic band (NEB) calculations and another based on MD simulations. Following the previously developed approaches for modal decomposition of thermal transport,[49-52] we project the atomic displacement or velocity fields onto the normal modes of vibration along the hopping trajectory obtained from a NEB calculation. The magnitude of projection determines which phonons/normal modes become excited to facilitate that specific ion hop. Our results suggest that only a small subset of phonons is responsible for ion diffusion in Ge-substituted $Li_3PO_4$ and that diffusivity can be increased only by increasing the temperature of these modes instead of the bulk temperature.

## Methods

**Modal Decomposition Based on NEB calculations**

Ion diffusion in solids typically involves the translation of an ion from one location to another, where the starting and final locations usually correspond to ion configurations that are local minima in the potential energy surface (see Fig. 1). Consequently, each of the equilibrium ionic configurations at the beginning and end of a hop could have a different set of normal modes of vibration. Thus, although we may have one minimum energy pathway (MEP) between equilibrium configurations 1 and 2 (see Fig. 1), because of the different vibrational modes at these two energy minima, the modes that initiate an ion hop from site 1 to 2 may be different from the ones that initiate a hop from site 2 to 1. More explanation on how two different equilibrium configurations in one structure possess different normal modes of vibration is provided in the ESI. To determine which modes of vibration in a local equilibrium contribute to the ion hop along a specific pathway, we combine NEB and lattice dynamics (LD) calculations.

An NEB calculation [59-62] determines the MEP between two equilibrium configurations. Each NEB calculation results in a desired number of ionic configurations (snapshots of displaced ions) that are distributed along the MEP between the two end equilibrium configurations. According to the LD formalism,[17] any atomic displacement field can be projected onto any complete set of normal modes of vibration to determine how the normal modes superpose to recreate that exact displacement field. The magnitude of projection, as has been recently learned from modal decomposition of thermal transport,[49-52] shows the degree to which each normal mode is contributing to that displacement field, which can be quantified using the following expression,[17, 51]

$$Q_n = \sum_{i=1}^{N} \sqrt{m_i}\, \mathbf{e}^*_{i,n} \cdot \mathbf{u}_i \quad (1)$$

where $\mathbf{u}_i$ is the displacement of atom $i$ from its equilibrium position in the configuration at the hopping origin, $\mathbf{e}_{i,n}$ is the eigenvector for mode $n$ assigning the direction and displacement magnitude of atom $i$ obtained from a super-cell LD (SCLD) calculation[50, 51, 63] for the structure at the hopping origin, $*$ denotes the complex conjugate operator, and $m_i$ is the mass of atom $i$. $Q_n$ is the modal displacement coordinate, the square of which is proportional to the mode energy $E_n$ according to,

$$E_n = \frac{1}{2}\omega_n^2 Q_n^2 \quad (2)$$

The total energy of the system $E$ is equal to the summation over all modal energy values, $E = \sum_n E_n$. Here, $E_n$ is the contribution by mode $n$ to the energy of the dispalced lattice during the ion migration, which can be interpreted as the contribution by mode $n$ to the ion hop along its migration pathway. To perform the projection of the displaced atoms onto the normal modes, one must make a choice as to which configuration along the MEP should be used to perform the projection. Although this is somewhat arbitrary, for simplicity, we selected the configuration that is halfway between the hopping origin and the transition state (see Fig. 1). We avoided projection on points closer to the saddle point based on the practical consideration that the structure would likely exhibit modes with imaginary frequencies near the transition state, which might introduce undesirable complexity, the effect of which will be analyzed in follow on studies. Nevertheless, we examined the projections closer to the beginning stages of the MEP, compared them to the halfway point, and noticed almost no change in the contributing modes (Fig. S1), which suggests the choice of the halfway point is satisfactory, at least for the system under consideration. To calculate the modal contributions for all the possible hops in the structure, we applied the above procedure to (*i*) all the distinct equilibrium configurations in the system since each of them would have different vibrational modes, and (*ii*) all the ion hops in each of these equilibrium configurations.

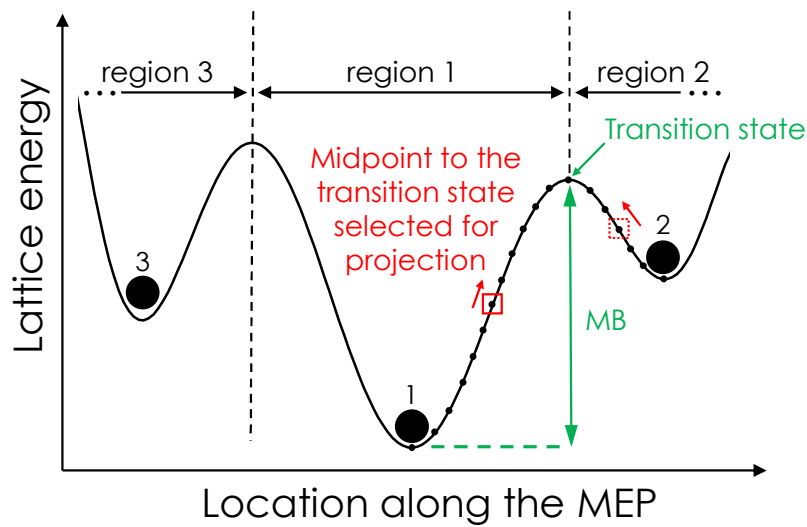

**Figure 1.** MEP schematic between three minimum energy sites 1, 2, and 3. Migration barrier (MB) is the maximum energy difference along the MEP (e.g., MB=$E_{\text{transition state}}$–$E_1$ for the hop from site 1 to 2) that the hopping ion needs to overcome. The dots on the MEP between 1 and 2 show the ionic configurations (displacement fields) obtained from the NEB calculation. The displacement fields halfway point to the transition state shown by the solid and dashed squares are used to obtain the modal contributions to the ion hops from sites 1 to 2 and 2 to 1, respectively.

To test our formulations, we chose Ge-substituted $Li_3PO_4$ as an example structure. The interatomic potentials for NEB calculations and MD simulations were obtained from a well-established form potential that has been successfully tested in a number of previous studies,[64-68] including Li-conducting compounds.[66, 68] This form potential is comprised of a long-range Coulomb term, a short-range Morse function, and a repulsive term. The parameters were obtained from the library of potentials developed by Pedone et al.,[69] which have been successfully tested in previous MD simulations of silicates and polyanion type materials including $Li_3PO_4$.[67] The exact formulas and parameters of which are provided in Table S1. To further check the accuracy of this utilized interatomic potential, we calculated the lattice constants for the perfect γ-$Li_3PO_4$ crystal from the isobaric-isothermal MD relaxation at zero pressure and room-temperature and compared them with the existing first-principles[70] and experimental[71-73] values (Table S4). The comparison showed that the lattice constants in our simulations have less than 3% difference with other reported values. The pristine $Li_3PO_4$ structure was chosen to be a 3×2×1 supercell containing 192 atoms, (γ-$Li_3PO_4$, space group *Pnma*) with lattice dimensions given is Table S4. In the pristine structure, one P was substituted by one Ge and one Li was added in the form of an interstitial to maintain charge neutrality, hence the structure contains 193 ions ($Li_{3.042}Ge_{0.042}P_{0.958}O_4$). This composition is a simple case to show the effectiveness of the modal excitation approach in this manuscript. If the number of Ge substitute atoms increase, for charge neutrality, the number of Li interstitials should also increase. More number of interstitial sites will complicate the modal excitation experiment because of the possibility of simultaneous hops in the simulation and the consequent challenge in deciding what modes to excite during the simulation. There are 36 available unique interstitial sites to add this Li interstitial. However, among these 36 interstitial sites, only 31 resulted in a stable configuration with positive vibrational frequencies (Fig. S2 shows schematics of these distinct configurations). Each hopping event brings the structure from one of these equilibrium configurations to another.

As is shown in the results section, the dominant hopping mechanism in our Ge-substituted $Li_3PO_4$ is the interstitialcy mechanism.[74-77] To determine the phonon contributions to all the possible interstitialcy hops in the system, we need to account for all the interstitialcy hops in each of the equilibrium configurations. To do so, we first counted the nearest neighbor Li lattice sites to the interstitial ion. Then, for each of these nearest neighbor Li lattice sites (tetrahedral sites), we counted the nearest neighbor interstitial sites (octahedral sites), with the constraint that the nearest neighbor interstitial site be among the 31 stable interstitial sites. Counting for all these combinations, results in 619 distinct hops in our structure that follow the interstitialcy mechanism. The SCLD calculations determined the normal modes of vibration for all the 31 equilibrium configurations and were performed using the General Utility Lattice Program (GULP)[78] package, while the MD simulations and the NEB calculations were conducted using the Large-scale Atomic/Molecular Massively Parallel Simulator (LAMMPS)[79] package following the formulations by Henkelman *et al.*[60] to ensure the correct determination of the saddle point. The reasons for why our calculations are based on classical interatomic potentials and LAMMPS package are discussed in the ESI.

**Targeted Excitation of Modes to Enhance Diffusivity**

After finding the strongest contributing modes to the ion diffusion using the NEB approach, we examined if external excitation of a small subset of these modes in an MD simulation had any measurable impact on the tracer and conductivity ion diffusivities. At each time step during the simulation, the equilibrium configuration was found by checking where the interstitial ion was located (i.e., which octahedral site was occupied), as the distinction between equilibrium configurations was only based on the occupied interstitial site, and as the relative position of non-interstitial ions remained unchanged across different equilibrium configurations. By knowing the equilibrium configuration, we then excited the 5 highest contributing modes among all the contributing modes to all the possible hops belonging to that configuration, while keeping the bulk temperature fixed. After a hop happened, a new equilibrium configuration was detected, and the process continued by exciting the 5 highest contributing modes among all the possible hops associated with the new configuration.

Since ion diffusivity increases with temperature, we ensured that the bulk lattice temperature remained constant at low temperature, so that any change in the diffusivity could only be attributed to the excitation of the targeted modes and not the bulk temperature. To do so, we kept the total kinetic energy of the system constant via a velocity rescaling scheme, whereby the addition of energy to the top 5 modes was complimented by a uniform reduction in the kinetic energy of all other modes in the system. To change the temperature of mode $n$ to a desired temperature $T_d$, we modified the atomic velocities in the system according to the following formula,

$$\mathbf{v}_i = \mathbf{v}_i + \frac{1}{\sqrt{m_i}} \left[ \sqrt{2 k_B T_d} - \dot{Q}_n(t) \right] \mathbf{e}_{i,n} \tag{3}$$

where $\mathbf{v}_i$ is the velocity of atom $i$, and $\dot{Q}_n$ is the modal velocity coordinate defined by,[17]

$$\dot{Q}_n = \sum_{i=1}^{N} \sqrt{m_i} \, \mathbf{e}^*_{i,n} \cdot \mathbf{v}_i \tag{4}$$

The derivation of Eq. 3 is provided in the ESI. This rescaling procedure induces only a slight perturbation to the system given that the top 5 modes only comprise 0.86% of the modes in the system, and thus the reduction in energy for all other modes is only <1% of their energy (see the ESI for discussions about the needed thermalization power). In this study, the above mentioned rescaling procedure was applied every 5 times steps (every 5 fs).

**Modal Decomposition Based on MD simulations**

To directly obtain the contributing modes to the ion diffusion from MD simulations, we used the definition of diffusivity based on the fluctuation-dissipation theorem[80, 81] and followed a modal decomposition approach similar to that employed in MD-based thermal transport studies.[49-52] Consistent with prior work on phonon transport,[49-52] we have termed this formalism Mass Diffusivity Modal Analysis (MDMA). In MDMA, the modal contributions to the atomic velocities are obtained and then used to decompose the following definition for conductivity diffusivity $(D_\sigma)$,[81-85]

$$D_\sigma = \frac{N_c}{3} \int_0^\infty \langle \mathbf{v}_c(\tau) \mathbf{v}_c(0) \rangle d\tau \qquad (5)$$

where $N_c$ is the number of hopping particles in the system, $\langle \cdots \rangle$ denotes the auto-correlation operator, and $\mathbf{v}_c$ is the center of mass velocity for the hopping particles,

$$\mathbf{v}_c = \frac{1}{N_c} \sum_{i=1}^{N_c} \mathbf{v}_i \qquad (6)$$

To calculate the individual contribution by mode of vibration $n$ to $\mathbf{v}_i$ $(\mathbf{v}_{i,n})$, first modal velocity coordinate $\dot{Q}_n$ is calculated (Eq. 4),[17] from which $\mathbf{v}_{i,n}$ is determined by,[17]

$$\mathbf{v}_i = \sum_{n=1}^{3N} \frac{\dot{Q}_n(t)}{\sqrt{m_i}} \mathbf{e}_{i,n} = \sum_{n=1}^{3N} \mathbf{v}_{i,n}$$

$$\rightarrow \mathbf{v}_{i,n} = \frac{\dot{Q}_n(t)}{\sqrt{m_i}} \mathbf{e}_{i,n} \qquad (7)$$

Replacing $\mathbf{v}_i$ in Eq. 6 with its modal contributions $(\mathbf{v}_{i,n})$ results in the modal contributions to the carriers' center of mass,

$$\mathbf{v}_c = \sum_{n=1}^{3N} \frac{1}{N_c} \sum_{i=1}^{N_c} \mathbf{v}_{i,n} \qquad (8)$$

which is substituted in one of the $\mathbf{v}_c$ terms in Eq. 5 to extract the modal contributions to diffusivity $(D_{\sigma,n})$,

$$D_\sigma = \sum_{n=1}^{3N} \frac{1}{3} \int_0^\infty \left\langle \sum_{i=1}^{N_c} \mathbf{v}_{i,n}(\tau) \mathbf{v}_c(0) \right\rangle d\tau = \sum_{n=1}^{3N} D_{\sigma,n}$$

$$\rightarrow D_{\sigma,n} = \frac{1}{3}\int_0^\infty \left\langle \sum_{i=1}^{N_c} \mathbf{v}_{i,n}(\tau) \mathbf{v}_c(0) \right\rangle d\tau \tag{9}$$

Similarly, the modal contributions to ion conductivity $(\sigma_n)$ can be obtained, the details of which is provided in the ESI.

By substituting for both center of mass velocities in Eq. 5 with their modal contributions from Eq. 7, the above formulations can be extended to capture the modal contributions to diffusivity in more detail,

$$D_\sigma = \sum_{n=1}^{3N}\sum_{n'=1}^{3N} \frac{1}{3}\int_0^\infty \left\langle \sum_{i=1}^{N_c} \mathbf{v}_{i,n}(\tau) \sum_{i'=1}^{N_c} \mathbf{v}_{i',n'}(0) \right\rangle d\tau = \sum_{n=1}^{3N}\sum_{n'=1}^{3N} D_{\sigma,n,n'} \tag{10}$$

where individual contributions from the correlation/interaction of eigen mode pairs $n$ and $n'$ $(D_{\sigma,n,n'})$ are given by,

$$D_{\sigma,n,n'} = \frac{1}{3}\int_0^\infty \left\langle \sum_{i=1}^{N_c} \mathbf{v}_{i,n}(\tau) \sum_{i'=1}^{N_c} \mathbf{v}_{i',n'}(0) \right\rangle d\tau \tag{11}$$

which provides additional insight into the degree to which each pair of modes interact, i.e. anharmonicity via terms where $n \neq n'$, to facilitate diffusion.

The combination of the MD and NEB based modal decomposition methods presented herein provide a detailed picture for the ion diffusion. Although both methods are based on projecting the trajectory of the hopping ion on the eigenvectors of vibration, the way they sample the phonon contributions to the ion hop is different. In the NEB-based approach the contributions are calculated based on the atomic displacement field, while in the MD-based method, the contributions are obtained based on the atomic velocity field. In addition, MD samples the dominant mechanisms that are present in a natural simulation of the material, while NEB can be used to study specific mechanisms and ion hop along specific migration pathways, regardless of the likelihood that they would occur at a given temperature.

**Diffusivity calculations**

Total mean squared displacment method (TMSD) was utilized to measure the diffusivity of Li ions in our structure, the details of which is explained in the existing literature[86, 87] and in the ESI. First, the structure is relaxed under the isobaric-isothermal ensemble for 1 ns at zero pressure and T=400 K and under the Nose-Hoover canonical ensemble for another 1 ns at T=400 K, then the structure was simulated for another 10 ns under the same ensemble to obtain the needed dispalcement data for diffusivity calcualtions. The timestep was chosen to be 1 fs, and statistical uncertainty was reduced by considering 5 independent ensembles.[88] For calculating the conductivity diffusion coefficients, total conductivity was calculated following the existing literarure,[84] the details of which are also explained in the ESI.

## Results and discussions

**Modal contributions obtained from NEB calculations of Ge-substituted $Li_3PO_4$**

The modal contributions to one example hop in Ge-substituted $Li_3PO_4$ shows that a small subset of modes (around 10 THz) contributed much greater than the rest of the modes (Fig. 2a). The dominant hopping mechanism in Ge-substituted $Li_3PO_4$ structure, is the concerted hop[89] of two Li ions, one of which is the interstitial, as shown in Fig. 3c and compared to two other hopping mechanisms in Figs. 3a and b. The concerted hop mechanism is known as the interstitialcy mechanism,[74-77] which has been reported[70, 90] as the dominant mechanism for lithium ion diffusion in $Li_3PO_4$ and is used as the migration mechanism for the NEB-based modal analyses presented in this study. The normalized integrations of the contributions with respect to frequency (the accumulation) are shown as a continuous curve in Fig. 2b for all the 619 hops in the structure. The majority of the contributions to all the hops in the system come from modes between 8 and 20 THz (Fig. 2b). Although this range is centered around the attempt frequency of lithium ion (~14 THz, see Fig. S3 and the ESI for the calculation procedure), the large range of frequencies for the contributing phonons shows that more intricate phonon processes are at play, which differs from the simpler picture described by traditional approaches,[91] wherein only one value of vibrational frequency (attempt frequency) enters the formulation. By averaging over all the modal contributions calculated for all the hops in the system (Figs. S4), we confirmed that on average >87% of the lattice energy during the ion hop in the structure came from <10% of the modes in the system (Fig. S5). In addition, as is explained and shown in Fig. S11, in each hopping event, there is at least one mode that has a contribution more than 10 times the average contribution by all the modes. Thus, the results in Fig. 2 show that in Ge-substituted $Li_3PO_4$, only a small number of modes are responsible for the interstitialcy diffusion mechanism in Fig. 3c.

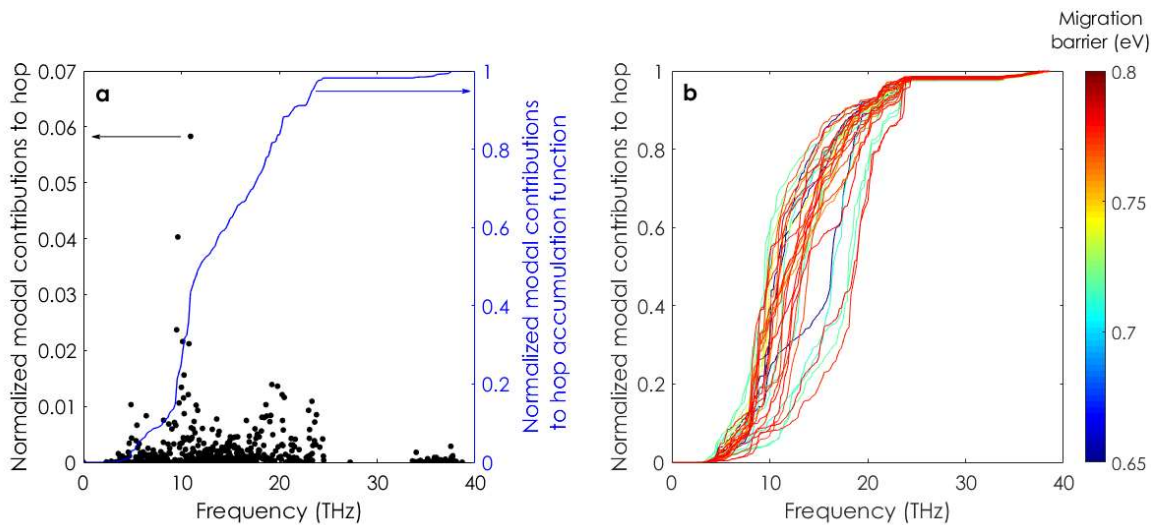

**Figure 2.** Modal contributions to the ion hops in our model Ge-substituted Li$_3$PO$_4$ structure obtained from Eqns. 1 and 2. (**a**) Modal contributions in the form of scattered (black dots) and accumulation (blue line) plots for one individual Li ion hop – the energy of each mode is normalized by the lattice energy in that snapshot, and (**b**) modal contributions in the form of the accumulation plots for all the 619 detected hops in our model Ge-substituted Li$_3$PO$_4$ structure. Because of the broken symmetry, different pathways in the system will theoretically have different migration barriers, the distribution of which is shown by different colors in panel (**b**).

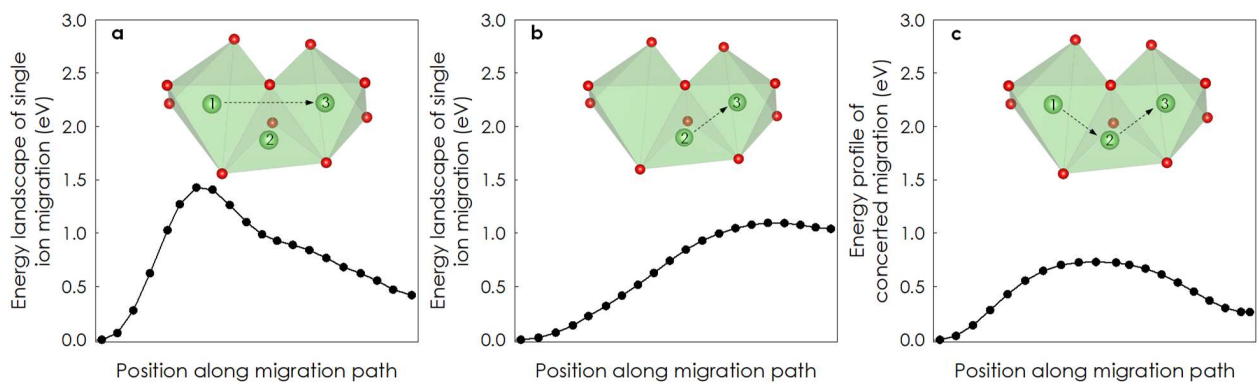

**Figure 3.** Three hopping mechanisms in Ge-substituted Li$_3$PO$_4$ structure and the energy along the MEP (dotted curves) obtained from NEB calculations: (**a**) the direct hop of a Li ion form an octahedral (interstitial) site (1) to another unoccupied octahedral site (3), (**b**) the hop of a Li ion from a tetrahedral (lattice) site (2) to an unoccupied octahedral site (3), and (**c**) the concerted hop of two Li ions, where one hops from a tetrahedral site (2) to an unoccupied octahedral site (3) and the other hops from an octahedral site (1) to the tetrahedral site that was previously occupied by the first Li ion (2). The dominant hopping mechanism in Ge-substituted Li$_3$PO$_4$ structure, is the concerted hop of two Li ions (panel (**c**)), because of its lower migration barrier than the other two mechanisms.

Due to the larger amplitudes of vibration that localized modes attribute to the interstitial and other neighboring ions than the delocalized modes[92] (Fig. 4), localized modes might be the most strongly contributing modes to each ion hop. However, inspection of the eigenvectors for the top two contributing modes in Fig. 2a (at 9.64 THz and 10.94 THz) revealed that highly contributing modes can be both delocalized (Fig. 4a) or localized (Fig. 4b) in nature. Since the modal analysis formalism is based on the SCLD calculations combined with Eqns. 1 and 2, all types of vibrational modes in the system with no inherent assumptions about their character were identified from which the contributions to ion diffusion even for localized modes were computed. Having access to such a formalism is crucial, because by breaking the symmetry in the system (e.g., by inclusion of vacancies, interstitial and substitutional ions, which is common in highly conductive solid electrolytes[93]) the conventional picture of propagating phonons in pure crystalline solids breaks down as other non-propagating and localized modes must be included.[54, 63, 94] Figure S6 shows the spectral distribution of all the localized modes of vibration for all the equilibrium configurations in our Ge-substituted $Li_3PO_4$.

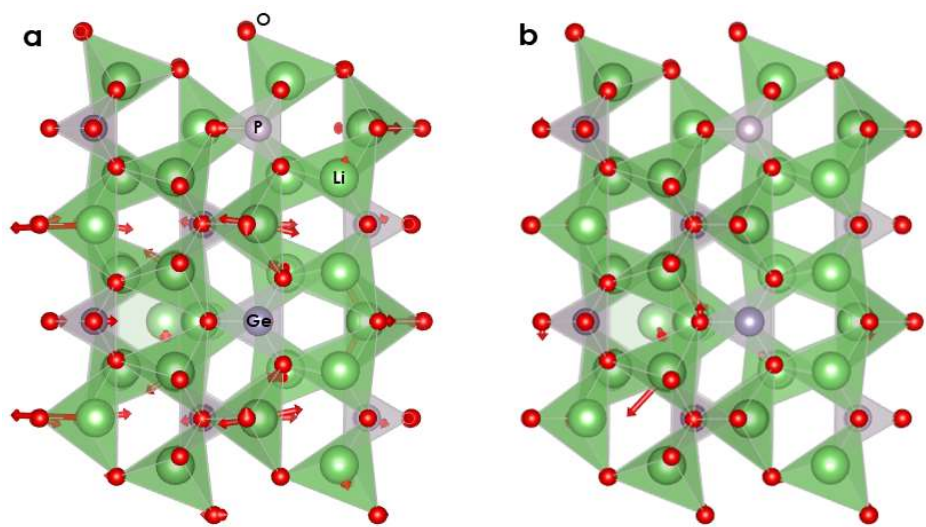

**Figure 4.** Eigenvectors of vibration (red arrows) for the (**a**) highest and (**b**) second highest contributing modes in Fig. 2a. Eigenvectors follow a (**a**) uniform and (**b**) non-uniform distribution, which correspond to the delocalized and localized nature of these depicted modes, respectively. The chosen equilibrium atomic configuration corresponds to panel 6 in Fig. S2.

**Diffusivity enhancement by excitation of the highly contributing modes**

Figure 5 shows that increasing the energy of the top 5 contributing modes among all the possible hops belonging to the equilibrium configuration present at each time step during an MD simulation, by increasing their temperature to 500 K and 700 K, increases the diffusivity by 2 (~$10^{-10}$ cm$^2$/s) and 4 (~$10^{-8}$ cm$^2$/s) orders of magnitude respectively, compared to its value (~$10^{-12}$ cm$^2$/s) in the unperturbed (natural) simulation at a lower temperature of 400 K. In fact, when the lattice is kept at 400 K and only the top 5 most contributing modes are excited to a particular temperature, the diffusivity becomes what it would have been if the entire material were heated to that temperature. Unexpectedly, even if we only excited the single highest contributing mode in each of the equilibrium configurations, almost the same increase in the diffusivity was observed. The small difference between the increase in diffusivity by exciting the 5 highest contributing modes and the one by exciting the highest contributing mode can be explained by noting that more kinetic energy is input to the hopping ion by exciting 5 modes. How these excited modes interact with the rest of the unexcited modes in the system is an interesting topic for a future study, which can be investigated using the two-temperature model that was presented in a recent study.[18]

To illustrate that the increase in diffusivity due to excitation only happens when the specific modes that matter most are excited, Fig. 5 shows that exciting 5 random modes in the 8-20 THz frequency range did not lead to significant enhancement in the diffusivity. The novelty of discovering that only a small subset of vibrational modes are responsible for the orders of magnitude increase in diffusivity by modal excitation can be understood by noting the theoretical definition of diffusion coefficient in solid materials, $D = D_0 \exp\left(-E_A/k_B T\right)$.[91] In this definition, $E_A$ and $D_0$ are known as the activation energy and the prefactor. According to the transition state theory,[95] $E_A$ and the passing of ion through the transition state are solely influenced by one mode of vibration in the system. However, $D_0$ is theoretically influenced by many modes of vibration in the system, which can be understood by noting its dependence on several other parameters, $D_0 \propto a^2 n\nu \exp\left(\Delta S_m/k_B\right)$, where $a$, $n$, $\nu$, and $\Delta S_m$ are the lattice constant, number of the nearest hopping sites, average frequency of vibration for the hopping atom and the entropy of migration. Among these parameters, $\Delta S_m$ is a function of all the frequencies of vibration in the system and because of its exponential influence, it can affect the diffusion coefficient by orders of magnitude, to a degree comparable to the influence of migration barrier as has been shown in recent studies by Krauskopf et al.[44] and Muy et al.[68] Therefore, since theoretically many modes of vibration influence $\Delta S_m$ and hence $D_0$, the observation that by only exciting a few number of modes in the system the diffusivity can change by several orders of magnitude is indeed unexpected.

To check if traditional parameters for diffusivity[96] can explain the enhanced diffusion rates caused by modal excitation in Fig. 5, we quantified the changes in attempt frequency (Fig. S3), jump rate (Table S2), radial distribution function (RDF) (Fig. S7), amplitude of vibration (Fig. S8), Haven ratio (the ratio of the tracer diffusion coefficient to the conductivity (total) diffusion coefficient)[84] (Table S3), and the migration barrier (Fig. S10). However, the very small changes in these parameters were not able to explain the orders of magnitude increase in diffusivity observed by modal excitation (Fig. 5). This also highlights the novelty of the mechanism discovered herein, illustrating the importance of modal temperatures (i.e., modal energies) and the concept of directionality brought into the analysis by including the eigenvectors of vibration in our phonon-based methodologies. Overall, the 2-4 order of magnitude increase in diffusivity by modal excitation (Fig. 5) can be attributed to increasing the energy of the highly contributing modes in the structure, where these modes can push the ion forward along its migration pathway. These results not only show that the identified modes of vibration using our proposed methodologies are correct and are indeed responsible for the ion diffusivity, but also offer a new strategy to increase the diffusivity of known solid-state ionic conductors by using techniques such as direct[19-21] and indirect[23-25] coupling to phonons without increasing the temperature of the lattice.

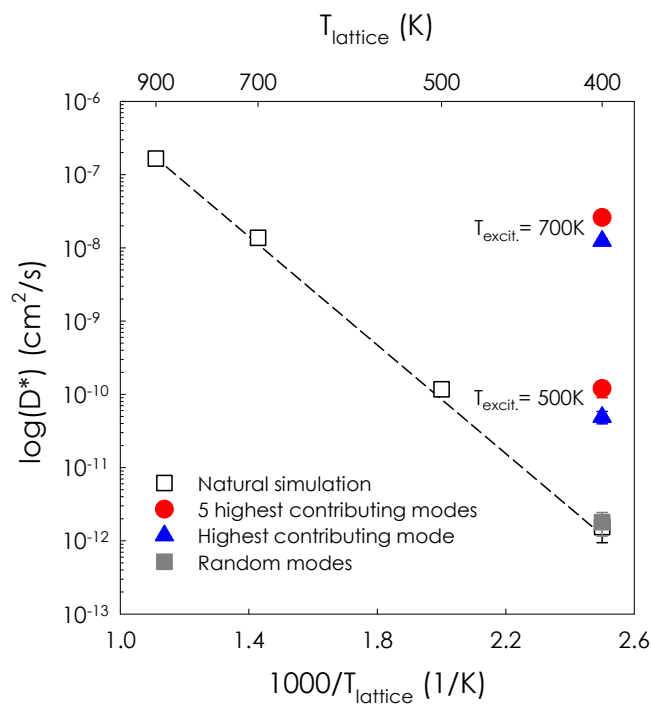

**Figure 5.** By exciting the 5 highest contributing modes or even the highest contributing mode to ion diffusion in the Ge-substituted Li$_3$PO$_4$ structure to 500 K or 700 K and keeping the lattice at 400 K, the diffusivity increases by around 2 and 4 orders of magnitude compared to the unperturbed/natural simulation at 400 K. Exciting 5 random modes to 700 K, however, does not lead to a noticeable increase in diffusivity. The reported values are tracer diffusivity, but the same increase was also observed in the conductivity (total) diffusivity[84] (see Table S3). The activation energy obtained from the linear fitting of the Arrhenius equation[86] (0.68eV) is in reasonable agreement with the one obtained from the NEB calculations for the interstitialcy hop (Fig. 3). In addition, the order of magnitude of our obtained diffusivity values are in agreement with the ones reported for similar structures in previous studies.[67, 68]

**Modal contributions obtained from MD simulations of Ge-substituted Li$_3$PO$_4$**

The modal contributions to the diffusivity for the Ge-substituted Li$_3$PO$_4$ structure, calculated from the MDMA method, are shown in Fig. 6b. The increase in the diffusivity accumulation function in Fig. 6b shows that the modes with frequencies between ~8-20 THz are responsible for ion diffusivity. This observation is in excellent agreement with the modal contributions obtained from the NEB-based approach (Fig. 2b), which is also averaged and shown in in Fig. 6b. The accumulation of the partial density of states (DOS) for lithium ions shown in Fig. 6b, obtained from the partial DOS in Fig. 6a, also shows reasonable agreement with the modal contributions obtained from the NEB-based and MDMA approaches. The generality of this agreement should be investigated more in future studies, however NEB and MDMA reveal the exact individual contributing modes to each ion hop in the system.

We use the MDMA formalism to, for the first time, quantify the anharmonic interactions that lead to ion diffusion in solid lattice. Individual values of $D_{\sigma,n,n'}$ (Eq. 11) are shown on maps of correlation of Figs. 6c and 6d. By summing over all the pairwise correlations on Fig. 6c, we confirmed that >92% of the contributions to diffusivity arise from the correlations along the diagonal (iso-frequency modes), indicating that the majority of the contributions by phonons to the diffusivity in the Ge-substituted Li$_3$PO$_4$ system originate from the harmonic interactions. The remaining contributions (~8%) came from off-diagonal terms (anharmonic interactions), which can be observed better in Fig. 6d after omitting the diagonal terms (i.e., artificially setting them to zero). Even after removing the diagonal terms, all the large contributions are still close to the diagonal, which shows that all the anharmonic terms are also between the modes that have relatively similar frequencies. Quantifying anharmonic interactions is particularly important for fast ion conductors, since the lower migration barriers in these ion conductors have been shown to be correlated with softer lattices,[43, 44] which typically allows for higher degrees of anharmonic interactions.[97] In a recent study, using AIMD simulations and Raman polarization-orientation measurements, Brenner et al.[98] attributed the large broadening of vibrational peaks in AgI superionic conductor to the strong anharmonic interaction of Ag$^+$ ion and the rigid iodine lattice. Our chosen Ge-substituted Li$_3$PO$_4$ compound is not a superionic conductor, which can be the reason for the low degrees of anharmonic contributions that we calculated for ion conduction from the MDMA method. Overall, our analysis provides a quantitative way to measure the effect of anharmonicity on the solid-state diffusion which has already been recognized in earlier works, but little in the way of quantitative evidence has been provided.[99, 100] This is with the exception of a few compounds, such as $CuCrSe_2$[101] and $AgCrSe_2$[102] where recently, inelastic neutron scattering measurements have revealed a softening of acoustic modes which is associated with the superionic transitions in these materials. Ultimately, more studies are needed to compare the inelastic interactions observed for our Ge-substituted $Li_3PO_4$ compound with other ion conductors with more polarizable anions where the anharmonicity are expected to be more important.[82, 103]

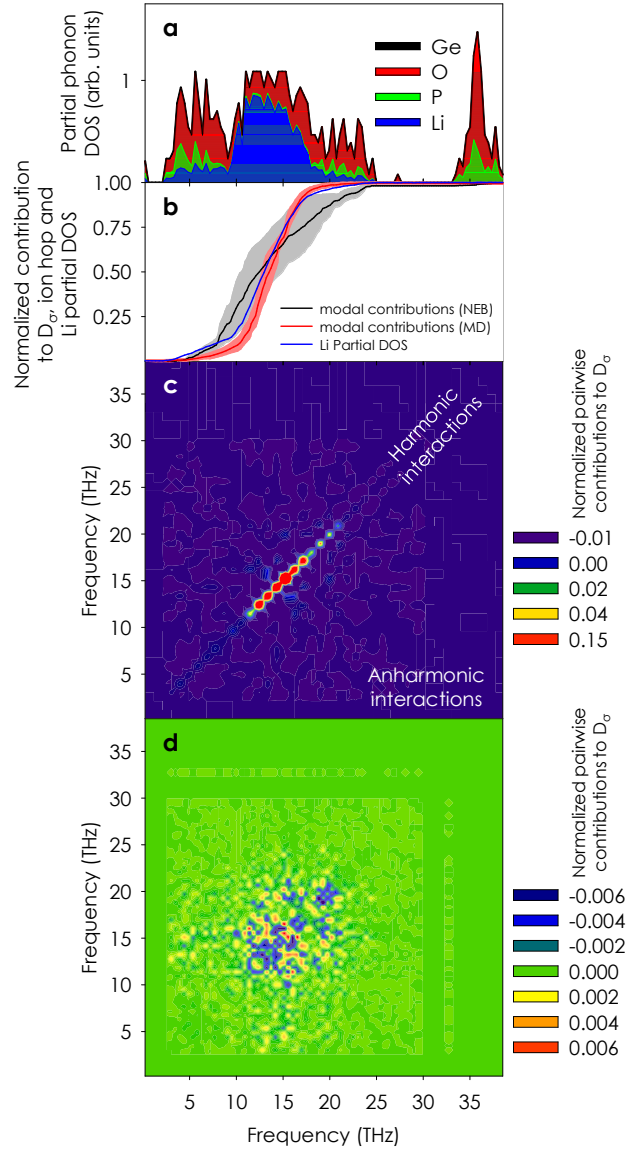

**Figure 6.** Normalized modal contributions to the diffusivity in Ge-substituted $Li_3PO_4$ structure calculated from MD based method (MDMA). The MDMA formalism is able to provide the degrees of both harmonic and anharmonic contributions to the ion diffusion in the lattice. (**a**) Partial phonon DOS for different elemental components of the structure in the form of stacked area chart. (**b**) Modal contributions in the form of accumulation function to the conductivity diffusivity obtained from the MDMA method (red curve) and to the ion hop from NEB based modal analysis (black curve), and the accumulation function for the lithium partial DOS (blue curve). (**c**) 2D map showing the magnitudes of the pairwise correlations/interactions contributing to the diffusivity on the plane of two frequency axes. (**d**) Similar to panel (**c**) except the values on the diagonal (harmonic interactions) have been zeroed for better visualization of the off-diagonal terms (anharmonic interactions). The shaded gray and red regions in panel (**b**) represent the statistical distributions in NEB (see Fig. 2b) and MD based modal decompositions, respectively. MD results shown on this plot are obtained from natural simulation at 800K.

## Conclusions

We have introduced two modal decomposition formalisms that determine the individual normal mode contributions to ion diffusion. Our results for a model structure of Ge-substituted $Li_3PO_4$ show that a small group of modes (<10% of the vibrational modes in the system) are responsible for >87% of the diffusion. Of greatest significance is that when we externally excited a targeted subset of the highest contributing modes, the diffusion increased to the value associated with the effective temperature of the targeted modes even if when the lattice is kept at lower temperature of 400 K. This observation is intriguing, since a high diffusivity can be achieved while the bulk temperature of the material remains at a much lower temperature.

While the selected Ge-substituted $Li_3PO_4$ composition in this study comprises a rather slow ion conductor, we believe that the presented rigorous theoretical frameworks in this report can be used to study the fundamental concept of phonon-ion interaction, which is important in the field of superionics.[104-107] Recent reports have so far only linked the idea that phonons contribute to ion transport,[33-44] but this is the first study showing that specific mode excitations can be used to influence diffusion in solids, which is influential to the field of lattice dynamics affecting ion transport and will most likely lead to more work on optoionics.[108-110] In addition, the idea that a diffusion hop can be enhanced directly by phonon excitation has far-reaching implications. If such an effect can be confirmed experimentally (see ESI for a possible approach to experimental evaluation of the reported observations in this study), it would open the door to many new possibilities. For example, it may be possible to operate systems or devices that have portions that are diffusion limited at lower temperatures, which could lead to major cost savings and performance improvements. For instance, hypothetically, if oxygen diffusion through YSZ were limiting in SOFCs, and the diffusivity it exhibits at a bulk temperature of 1100 K could be achieved at a much lower temperature of 300-400 K, via some other means, there might be a dramatic effect on lowering cost, enhancing efficiency, and increasing life and proliferation,[10] at the system level by using cheaper materials and seals for the piping/infrastructure. Furthermore, it is likely that this effect can be generalized to reactions, enabling a type of phonon catalysis,[111] whereby only the modes that matter are targeted and deliberately excited to accelerate a reaction. Similarly, this effect may be generalizable to phase transitions[31, 32] and other transformations, or possibly even the reverse i.e., the suppression of a reaction or phase transition by reducing the amplitude of the modes that facilitate the transition(s). The findings reported in this study therefore provide a gateway to thinking about new ways that specific vibrations can be used to enhance or potentially inhibit diffusion, reactions and/or phase transitions etc.

## Acknowledgements

We acknowledge support from National Science Foundation (NSF) Career Award to A.H. (Award No. 1554050), the Office of Naval Research (ONR) under a MURI program (Grant No. N00014-18-1-2429), and BMW.

## Author contributions

K.G., Y.S.-H, and A.H. conceived the idea. K.G. and A.H. derived the decomposition formulations. K.G. and S.M. performed the simulations. All authors contributed to the analyses, discussions and writing the manuscript.

## Conflicts of interest

There are no conflicts to declare.

## Supplementary Information

**Derivation of mode amplitude as a function of temperature**

At a fundamental level the total energy of the lattice $E$ is equal to the summation of energies from individual modes of vibrations $E_n$, where $E = \sum_n E_n$. The quantum mechanical description[17] of $E_n$ is given by,

$$E_n = \hbar\omega_n \frac{1}{e^{\frac{\hbar\omega_n}{k_B T}}-1} \tag{1}$$

where $n$ denotes mode index, $\hbar$ is the Planck's constant divided by $2\pi$, $\omega_n$ is the frequency of mode $n$, $k_B$ is the Boltzmann's constant and $T$ represents lattice temperature. If we can measure the energy of the mode $E_n$, we can use the following equation to calculate its temperatrue,

$$T = \frac{\hbar\omega_n}{k_B}\left[\ln\left(\left(\frac{\hbar\omega_n}{E_n}\right)+1\right)\right]^{-1} \tag{2}$$

At temperatures above the Debye temperature $(T > \theta_D)$, where the classical regime persists and is often the most relevant for technological applications, the definition of $E_n$ can be simplified to $E_n = k_B T$. Conisdering that each normal mode of vibration can act as a separate oscillator, it can have its own temparature, which we refer to it as modal temperature $T_n$. In addition, based on the classical equipartition theorem,[112] the potential energy of mode $n$ $(E_{n,P})$ will, on the other hand, be equal to half of its total energy. At the same time, potential energy of the mode is proportional to its modal vibrational amplitude $(Q_n)$ and frequency $(\omega_n)$, which results in,

$$E_{n,P} = \frac{1}{2}k_B T_n = \frac{1}{2}Q_n^2 \omega_n^2 \tag{3}$$

and the following expression for the mode amplitude,

$$Q_n = \frac{\sqrt{k_B T_n}}{\omega_n} \qquad (4)$$

**Calculation of vibrational modes at different local equilibrium configurations**

The different modes of vibration at different local equilibrium configurations can be understood and modeled mathematically by expanding the potential energy surface via a Taylor expansion[17, 113] about a given equilibrium configuration as,

$$E = E_0 + \sum_i \Pi_i u_i + \frac{1}{2!}\sum_{ij} \Phi_{ij} u_i u_j + \frac{1}{3!}\sum_{ijk} \Psi_{ijk} u_i u_j u_k + \cdots \qquad (5)$$

where $E$ represents the potential energy of all the interacting atoms, $E_0$ is the energy at the minimum, $u_i$, $u_j$, and $u_k$ are the displacements of atoms $i$, $j$, and $k$ from their equilibrium position at that particular equilibrium configuration, and $\Pi$, $\Phi$ and $\Psi$ are linear, harmonic and cubic force constants. In such an expression, the term including the linear force constants $(\Pi_i)$ diminishes at a stable equilibrium configuration because the atomic forces decay to zero. The second order terms $(\Phi_{ij})$ are then the largest non-zero terms, and most often the higher order terms, which affect the phonon-phonon interactions, can be safely neglected when we are simply interested in determining the mode shapes and frequencies. We can then solve the equations of motion for the entire system, via a super-cell lattice dynamics (SCLD) calculation. The output of such calculation is then a set of eigenvalues that describe the frequencies of each normal mode and a set of corresponding eigenvectors that describe the direction and magnitude of displacement that each atom in the system will experience when it participates in a specific normal mode (i.e. the mode shape). The modes are then specifically a result of the local equilibrium configuration, and in principle will differ for each distinct configuration.

**Derivation of the velocity rescaling formula**

After the equilibrium configuration is determined, we excite the modes based on the following formulation. To change the temperature of the normal mode of vibration $n$ to a desired tempearture $T_d$, we first calculate the modal velocity coordinate of that mode using Eq. 5. Using Eq. 6, $\dot{Q}_n$ is then used to calculate the modal contributions to atomic velocity $\mathbf{v}_i$ coming from the normal mode of vibration $n$ ($\mathbf{v}_{i,n}$) (Eq. 7 in the manuscript). Knowing that, for a classical system, the temperature of the normal mode $n$ can be calculated from $\frac{1}{2}\dot{Q}_n^2 = \frac{1}{2}k_B T$,[17] during the MD simulation, we modify the atomic velocities by first subtracting the effect of the modal components $\mathbf{v}_{i,n}$ from $\mathbf{v}_i$, and then adding the components that result in the desired modal temperature $T_d$ for mode $n$.

$$\mathbf{v}_i = \mathbf{v}_i - \frac{1}{\sqrt{m_i}}\dot{Q}_n(t)\mathbf{e}_{i,n} + \frac{1}{\sqrt{m_i}}\sqrt{k_B T_d}\mathbf{e}_{i,n} \quad (6)$$

which simplifies to,

$$\mathbf{v}_i = \mathbf{v}_i + \frac{1}{\sqrt{m_i}}\left[\sqrt{2k_B T_d} - \dot{Q}_n(t)\right]\mathbf{e}_{i,n} \quad (7)$$

**Modal contributions to ionic conductivity**

Ionic conductivity is proportional to diffusivity through the following relation,[82, 83]

$$\sigma = \frac{N_c(Z_c e)^2}{k_B T V}D_\sigma \quad (8)$$

where $Z_c$ is the charge of the carriers, $e$ is the electron charge, and $V$ is the volume of the system. After replacing the diffusivity by its modal contributions $D_{\sigma,n}$ from Eq. 9 in the manuscript, the above equation results in the following definition for modal contributions to ionic conductivity $\sigma_n$,

$$\sigma_n = \frac{N_c(Z_c e)^2}{k_B T V}D_{\sigma,n} \quad (9)$$

**Diffusivity Calculations**

Total mean squared displacement method (TMSD) is utilized to measure the diffusivity of Li atoms in our Ge-substituted Li$_3$PO$_4$. This method is well explained in existing literature,[86, 87] In short, following the Einstein relation, TMSD calculates the diffusivity of the hopping particles based on the slope of their mean square displacment (MSD) data over time interval $\Delta t$ as,

$$D = \frac{MSD(\Delta t)}{2d\Delta t} + D_{offset} \quad (10)$$

For three dimensional systems $d = 3$, $D_{offset}$ is the linear shift from the origin, and for tracer diffusivity, $MSD(\Delta t)$ is defined as,

$$MSD(\Delta t) = \frac{1}{N_{\Delta t}} \sum_{t=0}^{t_{tot}-\Delta t} \frac{1}{N} \sum_{i=1}^{N} |\mathbf{r}_i(t+\Delta t) - \mathbf{r}_i(t)|^2 \quad (11)$$

where $N$ is the number of hopping atoms/ions n the system, $N_{\Delta t}$ is the number of time inervals with the same duration $\Delta t$, $t_{tot}$ is the total simulation duration, and $\mathbf{r}_i(t)$ is the trajectory of the hopping atom/ion $i$. To obatin the data points in Fig. 5, we relaxed the structure at finite temperature and pressure under the isobaric-isothermal ensemble (NPT) for 1 ns at zero pressure and $T = 300K$ and under the canonical ensemble (NVT) for another 1 ns at $T = 300K$, after which we simulated the structure for another 10 ns under the NVT ensemble to obtain the needed displacement data for diffusivity calcualtions. A timestep of 1 fs was employed for all the simulations. Statistical uncertainty, due to insufficient phase space averaging, has been reduced by considering 5 independent ensembles.[88] The same exact simulation parameters were also utilized for modal excitation simulations with the only difference being that during the simulations the temperature of the desired normal modes in the system was changed to the desired temperature, as exaplined previously in the methods section. For calculating the conductivity diffusion coefficients used in the calculaiton of Haven ratios in Table S3, Eq. (10) is used in combination with the following formula, which is the MSD of the center of mass (COM) of Li$^+$ carriers,[84]

$$MSD_{COM}(\Delta t) = \frac{1}{N_{\Delta t}} \sum_{t=0}^{t_{tot}-\Delta t} \frac{1}{N} \left[ \sum_{i=1}^{N} |\mathbf{r}_i(t+\Delta t) - \mathbf{r}_i(t)| \right]^2 \qquad (12)$$

The curves shown on Fig. S9 suggest that the 10 ns simulation time that we had for this structure was suffcient to get convergence in our $D_\sigma$ simulations for Haven ratio calculations (Table S3).

**Calculations of Attempt Frequency, RDF, Amplitude of Vibration, and Jump Rate**

Attempt frequency (Fig. S3) is obtained from the weighted average of the partial phonon density of states of Li$^+$ ions, which is calculated from power spectrum of atomic velocities $v_{\alpha=x,y,z}$ extracted every 5 time steps (5fs), using[114, 115]

$$P_\alpha(\omega) = \left| \int v_\alpha(t) e^{-i\omega t} dt \right|^2 \qquad (13)$$

The total power spctrum is defined by $P_{total} = P_x + P_y + P_z$. RDF (Fig. S7) is calcualted using the included compute feature in LAMMPS[79] package. For capturing the overall structural position of atoms with respect to each other, the averaging was done over all pair possibilities in the system. Amplitude of vibration (Fig. S8) is obtained by weigthed averaging of the MSD of Li$^+$ ions in the system during a 200 ps simulation. Jump rate (Table S2) is found following the procedure explained by Klerk *et al.*,[96] where the number of jumps are counted in a known period of simulation time. To detect the occurnace of a jump, the LAMMPS[79] code was modified so it follows which interstitial site is occupied at each time step. A jump happens when the occuiped interstitital site changes, which is detected and counted during the simulation.[96]

**The reasons for choosing empirical interatomic potentials and LAMMPS package for NEB calculations and MD simulations**

As mentioned in the manuscript, empirical interatomic potential with Pedone's parameters[69] are used for NEB calculations and MD simulations using LAMMPS package[79]. The main reasons that the NEB and MD results presented in this manuscript were obtained from classical interatomic potentials and LAMMPS package are:

1. The MD-based modal decomposition approach presented in the manuscript is based on the conductivity diffusion formula that is based on the fluctuations-dissipation theorem.[84, 85] Previous studies have noted that compared to the mean-square displacement approach,[86, 87] the fluctuation dissipation approach[85] used in this study requires longer simulation times for convergence, which would be unachievable using AIMD approaches.[116] Finding a modal decomposition method based on direct mean square displacement (MSD) formulation[86] for diffusivity can help with faster convergence of the modal contributions. However, modal decomposition of the direct MSD formulation of diffusivity seems impossible because of the dependence of the atomic displacement on the reference position at the beginning of the simulation.

2. As explained in previous sections, to examine the effect of the selective modal excitation on the diffusivity, the atomic velocities in the system need to be modified, following the modal atomic velocity rescaling formula (Eqns. 6 and 7 in the ESI). Such a modification to the atomic velocities is achieved by modifying the LAMMPS velocity rescaling subroutine. Modifying the LAMMPS code is more straightforawrd than the other *ab-initio* codes. This when accompanied by the fact that longer simulation times are needed for modal analysis of lower temperatures such as 400K made us to utilize the LAMMPS package based on empirical interatomic potential.

3. In addition, as explained in the manuscript, 619 NEB and 31 supercell lattice dynamics were performed. For better analysis of the NEB results to capture the modal contributions to the ion hop, for each of the NEB calculations 17 images were used along the MEP which also adds to the computational expense.

Even if we are using a 193-atoms unit cell, the extent of these calculations and simulations makes *ab initio* approaches almost impossible. However, nothing restricts the introduced methodologies in this manuscript to be applied to first-principles methods

**Possible approach for the experimental evaluation of the reported observations**

A possible approach for experimentally evaluating the reported observations in this study is by illuminating the sample by THz photons[19-25] or THz electric fields.[26]. Ideally, the frequency of the illuminated photons or the oscillation of the electric field should be close to 10 THz, which corresponds to the sharp increase in the modal accumulation function in Fig. 2a in the manuscript. Any enhanced signal caused by the external stimuli can then be detected and compared to the baseline (no shining) performance using techniques such as THz time domain spectroscopy (THz-TDS) or Electrochemical Impedance Spectroscopy (EIS), which would confirm the validity of the reported observations.

**Table S1.** Interatomic potential parameters used for MD and NEB calculations. The interatomic potentials are described by the equation: $U(r) = \frac{z_i z_j e^2}{r} + D_{ij}\left[\left\{1-e^{-\alpha_{ij} \times (r-r_0)^2}\right\}-1\right] + \frac{C_{ij}}{r^{12}}$, where $r$ is the distance between two interacting atoms $i$ and $j$, $z$ represents the particle charge (also shown in the fist column of the table), and $e$ is the electron charge. The global cutoff for these potentials have been considered to be $15 A$.

| | $D_{ij}(eV)$ | $\alpha_{ij}(A^{-2})$ | $r_0(A)$ | $C_{ij}(eVA^{12})$ |
|---|---|---|---|---|
| $Li^{0.6} - O^{-1.2}$ | 0.001114 | 3.429506 | 2.681360 | 1.0 |
| $Ge^{2.4} - O^{-1.2}$ | 0.158118 | 2.294230 | 2.261313 | 5.0 |
| $P^{0.6} - O^{-1.2}$ | 0.831326 | 2.585833 | 1.800790 | 1.0 |
| $O^{-1.2} - O^{-1.2}$ | 0.042395 | 1.379316 | 3.618701 | 22.0 |

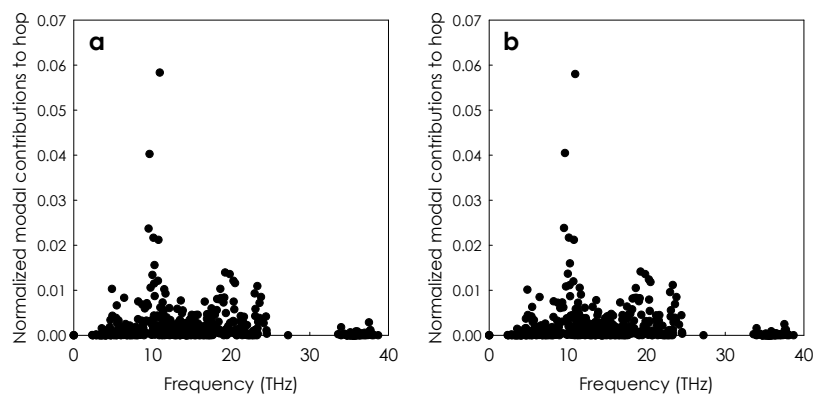

**Figure S1.** Choosing different points along the MEP for projection will result in almost unnoticeable change for the calculated contributing modes to the hopping process. Panel (**a**) corresponds to the midpoint to the transition state, which was used for all the calculation in the manuscript and shown with solid and dotted squares in Fig. 1, and panel (**b**) corresponds to the point in the first image in a NEB calculation (see Fig. 1). The calculated excitation energies for each mode in the two cases would be different, but the normalized energies would have similar behavior.

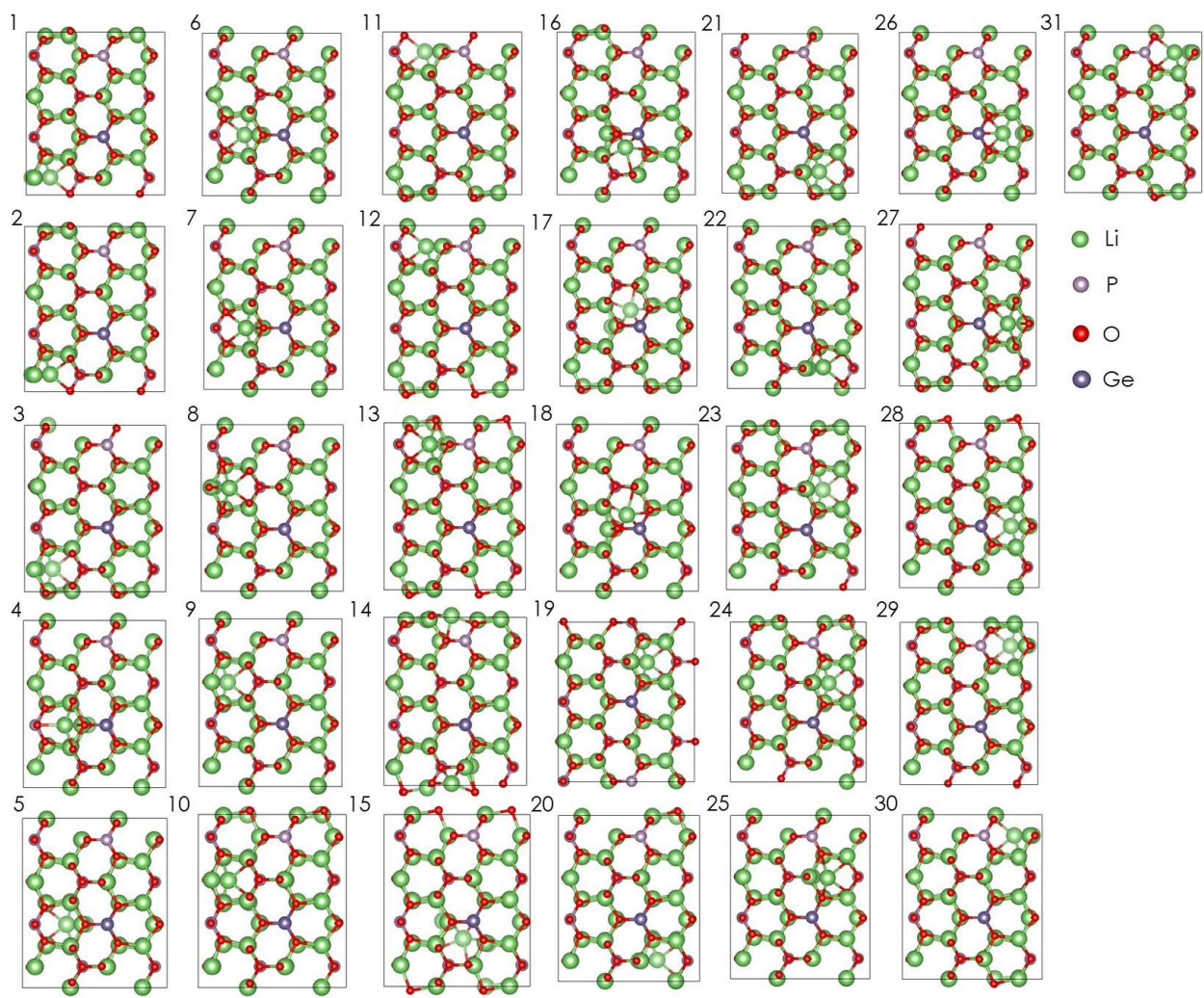

**Figure S2.** Schematic representation of the 31 equilibrium configurations in our Ge-substituted $Li_3PO_4$ structure. The type of the atoms in the structures are shown by colors noted on the right side of the figure.

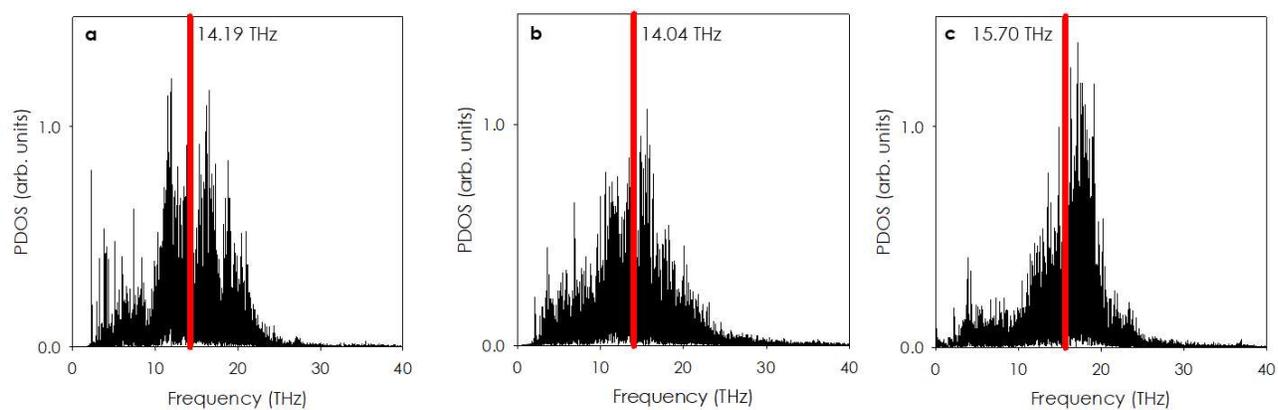

**Figure S3.** Attempt frequency shown by the red line,[96] which is the average of Li partial density of states (PDOS) (black line) obtained from the vibrational power spectrum of Li$^+$ ions during a MD simulation. The shown cases are for (**a**) natural vibrations at 400K, (**b**) natural vibrations at 700K, and (**c**) selective modal excitations at 700K, while the bulk is kept at 400K.

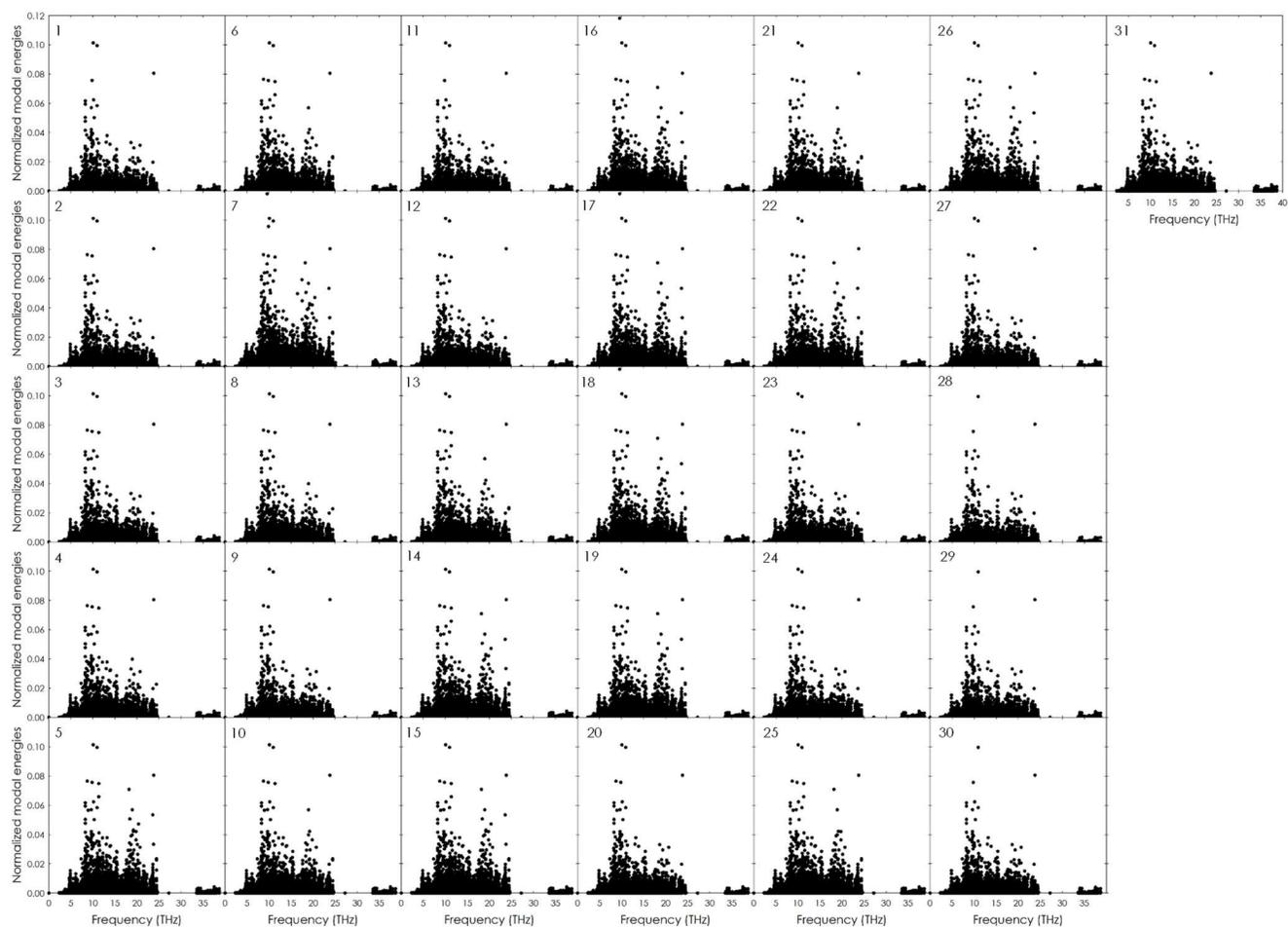

**Figure S4.** Modal contributions to all the 619 detected hops in our Ge-substituted $Li_3PO_4$ structure that follow the lowest activation energy hopping mechanism. Each panel includes the modal contributions to all the hops that have the same equilibrium configuration as their hopping origin. The panels above correspond to the 31 equilibrium configurations shown in Fig. S2.

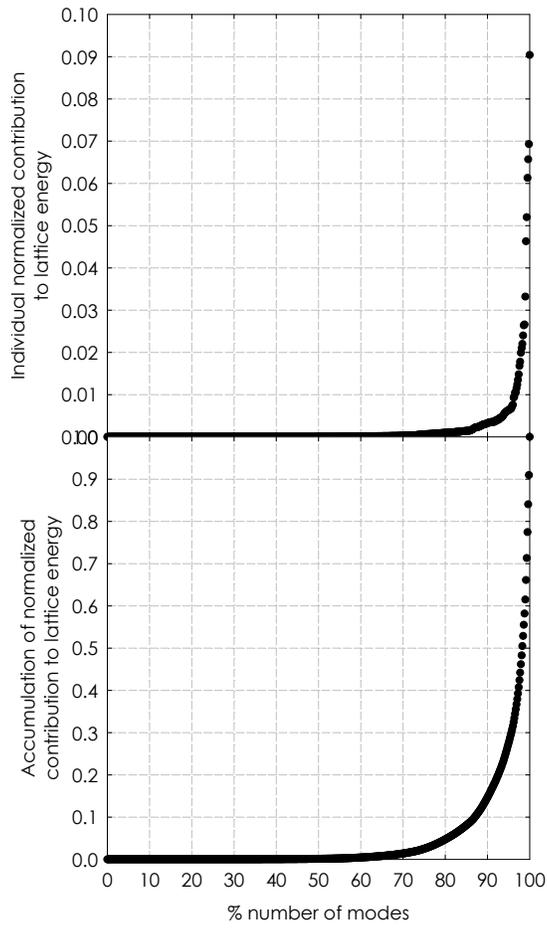

**Figure S5.** By changing the x-axis for each of the panels in Fig. S4 from frequency to percent number of modes, and averaging over all the modal contributions to the 619 hops in our system, which are shown in Fig. S4, and sorting the contributions based on their magnitude, it can be seen that more than 87% of lattice energy in the system during the hop comes from less than 10% of the modes.

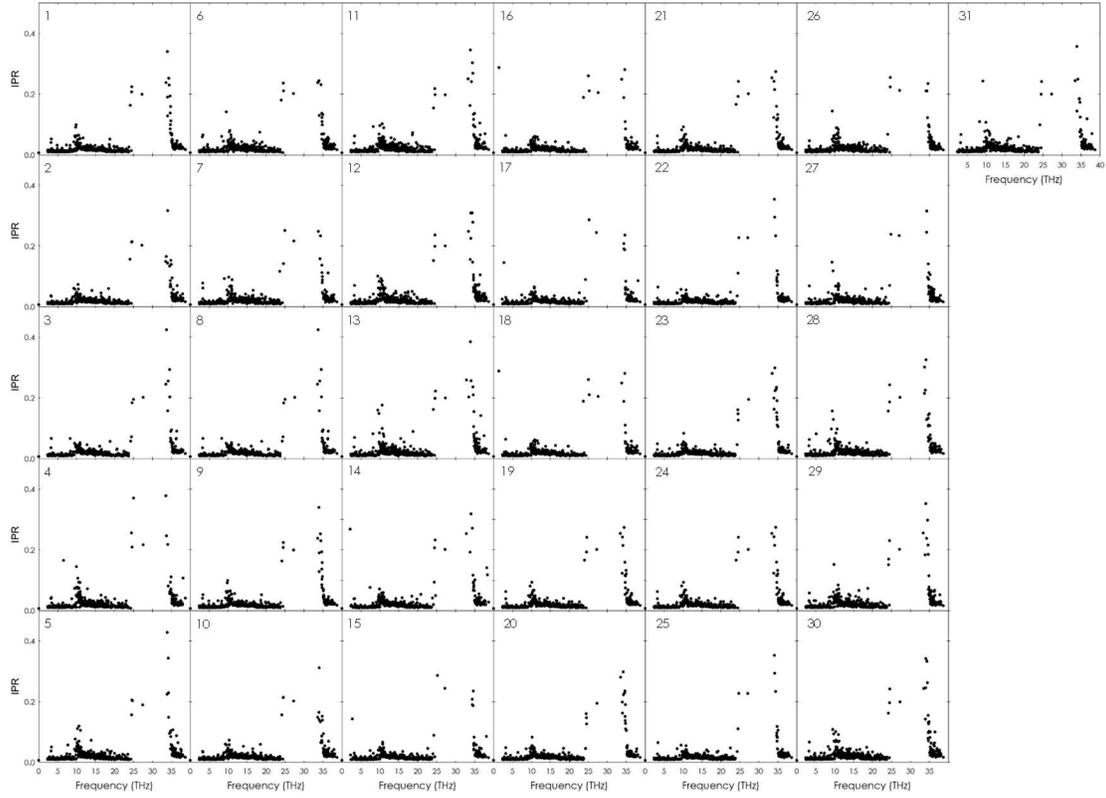

**Figure S6.** Inverse participation ratios (IPR)[117, 118] calculated for all the normal modes of vibration for the 31 equilibrium configurations in our Ge-substituted $Li_3PO_4$ structure. IPR is a direct index for localization of vibrational states in a system. A value of IPR ~ 1 corresponds to a highly localized mode, and a value of IPR ~ 0 (more accurately $1/N_a$ where $N_a$ is the number of atoms in the system) corresponds to a fully extended (i.e., delocalized) mode.[119] Panel numbers correspond to the equilibrium configurations shown in Fig. S2. By introducing even one Li interstitial to the system, localization happens, and considerable number of modes show IPR values much larger than 0. The formula used for IPR is as follows:

$$IPR = \left(\sum_i^{N_a} y_i^2\right)^2 \Big/ \sum_i^{N_a} y_i^4 \text{ ,}^{59,60}$$ where $y_i$ is the norm of eigenvector on each atom $i$ in the system (i.e., $y_i = \sqrt{(e_{i,x})^2 + (e_{i,y})^2 + (e_{i,z})^2}$, where $e_{i,x}$, $e_{i,y}$, and $e_{i,z}$ are the eigenvector components on atom $i$).

**Table S2**. Ionic jump rates for three cases of (**a**) natural vibrations at 400K, (**b**) natural vibrations at 700K, and (**c**) selective modal excitations at 700K, while the bulk is kept at 400K.

|  | 400K natural | 700K natural | 400K with targeted excitation to 700K |
|---|---|---|---|
| Jump rate (#jumps/ps) | 0.0007 | 0.138 | 0.241 |

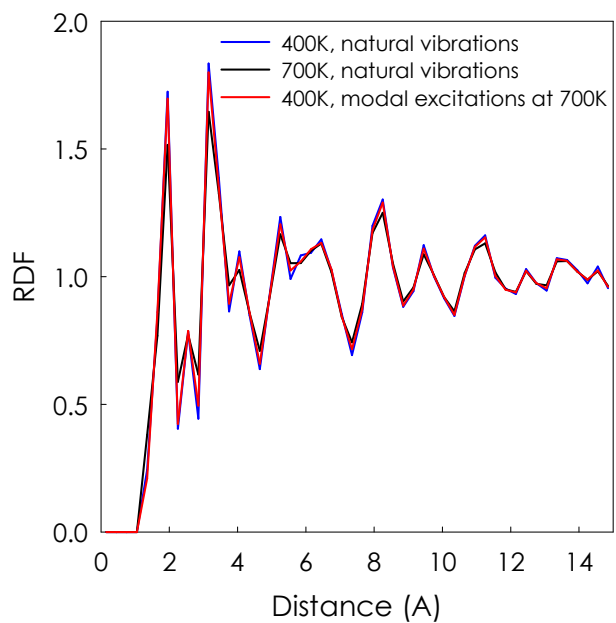

**Figure S7.** RDF for three cases of (**a**) natural vibrations at 400K (blue line), (**b**) natural vibrations at 700K (black line), and (**c**) selective modal excitations at 700K, while the bulk is kept at 400K (red line).

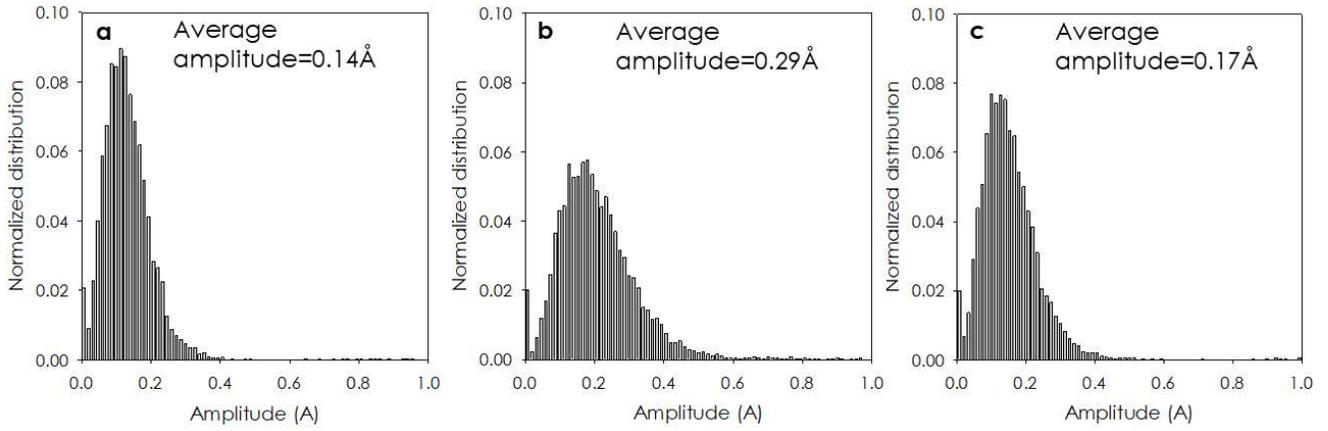

**Figure S8.** Distribution for amplitudes of vibration for three cases of (**a**) natural vibrations at 400K, (**b**) natural vibrations at 700K, and (**c**) selective modal excitations at 700K, while the bulk is kept at 400K.

**Table S3**. Change in tracer vs. conductivity diffusion and Haven ratio ($H_R$) for three cases of (**a**) natural vibrations at 400K, (**b**) natural vibrations at 700K, and (**c**) selective modal excitations at 700K while the bulk is kept at 400K. The change in the Haven ratio from ~0.8 without phonon excitation to ~0.4 with phonon excitation can be attributed to the fact that excited phonons contribute to the concerted hops in the system, which decreases the Haven ratio.[89]

|  | 400K natural | 700K natural | 400K, modes excited to 700K |
|---|---|---|---|
| $D^*\left(cm^2/s\right)$ | 8.17×10$^{-15}$ | 1.47×10$^{-8}$ | 4.11×10$^{-8}$ |
| $D_\sigma\left(cm^2/s\right)$ | 9.39×10$^{-15}$ | 2.07×10$^{-8}$ | 9.74×10$^{-8}$ |
| $H_R\left(D^*/D_\sigma\right)$ | 0.87 | 0.71 | 0.42 |

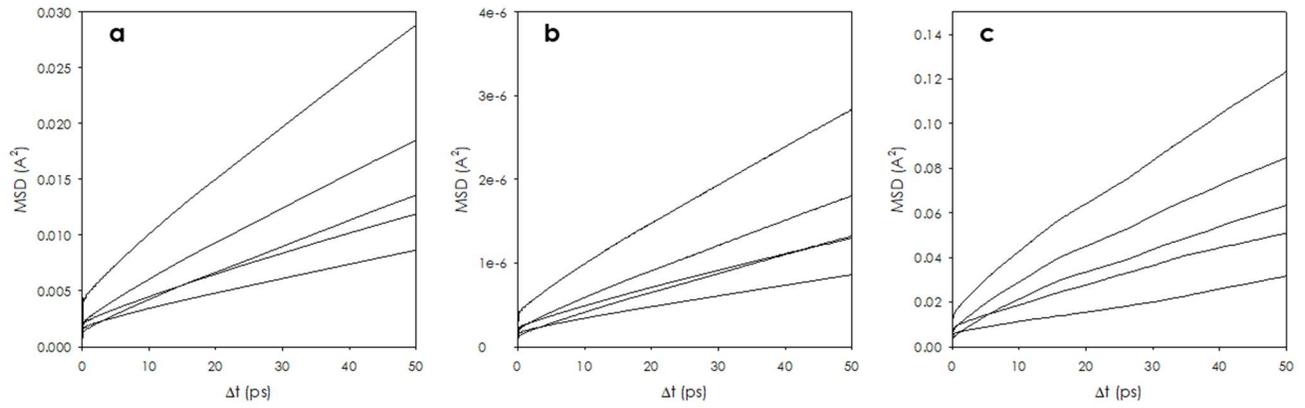

**Figure S9.** MSD of the carriers' (Li$^+$ ions') center of mass vs. Δt curves obtained from D$_\sigma$ calculations in Eq. 12 obtained from different ensembles for (**a**) a natural simulation at 700K, (**b**), a natural simulation at 400K and (**c**) an excitation simulation where the bulk is at 400K and selected modes are excited to 700K.

**Table S4.** Comparison of lattice constants for perfect γ-Li$_3$PO$_4$ obtained from the isobaric-isothermal MD relaxation at zero pressure and room-temperature using the Pedone potential parameters[69] with the existing first-principles[70] and experimental[71-73] values.

|  | First-principles[70] | Experiment[71] | Experiment[72] | Experiment[73] | This work |
|---|---|---|---|---|---|
| $a\,(Å)$ | 10.58 | 10.53 | 10.46 | 10.49 | 10.35 |
| $b\,(Å)$ | 6.17 | 6.12 | 6.11 | 6.12 | 6.01 |
| $c\,(Å)$ | 4.99 | 4.93 | 4.92 | 4.93 | 4.90 |

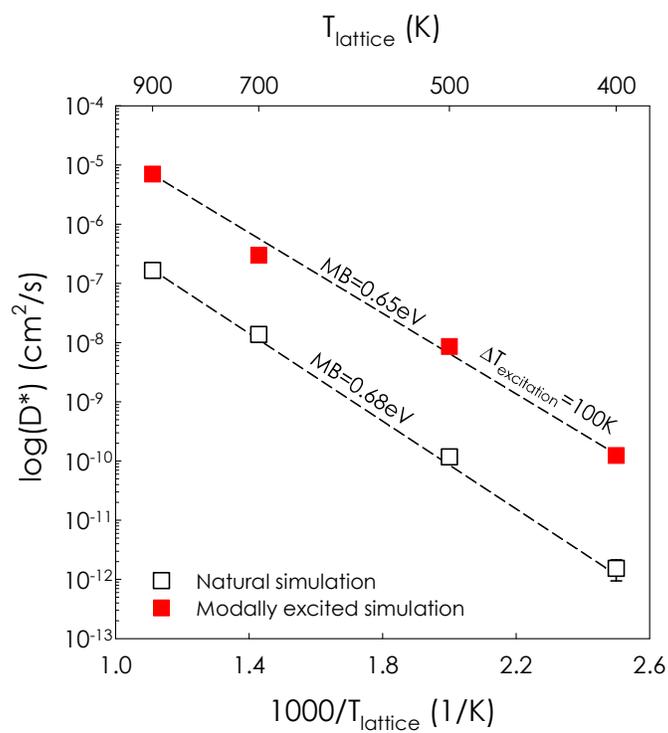

**Figure S10.** Diffusivity values at different lattice temperatures obtained from natural and modally excited simulations. In the modally excited simulations, the excitations increase the temperature of the top 5 contributing modes to 100K above the lattice temperature. It can be seen that the change in the activation energy between the natural and modally excited simulations are small and cannot justify the orders of magnitude increase in diffusivity that is caused by modal excitations.

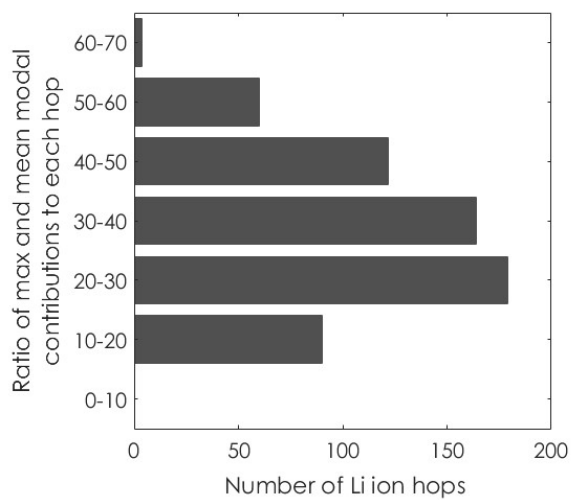

**Figure S11.** Histogram of the ratio of maximum to mean modal contributions for all the 619 ion hops in Ge-substituted $Li_3PO_4$ structure. Each hop was separately analyzed to generate a scatter plot like the example in Fig. 2a. The purpose was to examine if each hop showed the same type of behavior, where there was a small subset of modes with much greater amplitude than the rest of the modes. For each hop, we quantified the ratio of the maximum modal contribution to the average over all modes, which provides a clearer picture as to whether there is a dominant mode in every hopping instance. The histogram of the distribution of this ratio for all possible hops shows that in every case, there is at least one mode with a contribution >10X the average for all modes.

**Assessing the thermalization power and excitation rate needed for the modal excitation simulations**

The power needed to keep the lattice temperature at 400K is shown in Table S5. It can be seen that the power needed for thermalization for our 192 atoms system is impractically large. However, this issue is an artifact of the small size of our simulation cell, and the power deposited for modal excitation does not scale with the system size, because the energy is deposited in a local manner. What matters is that the highly contributing modes are excited and this amount of power does not need to be dumped in such a small volume. For example, exciting modes to 700K is key to getting a diffusivity value comparable to the one that is obtained at a 700K bulk temperature, but the high power density is not necessary, because the energy input is local and thus need not scale linearly with the system size.

This issue can be understood better if we consider again the simulations that were done by exciting the highest single contributing mode in the system. When we increase the energy of the highest singly contributing mode, the energy is deposited in a very localized region in space, however as we increase the size of the system the required power density decreases accordingly (see Table S5). This decrease in power density can be justified by noting that if we increase the number of atoms we correspondingly decrease the size of the eigenvectors on the modes. However, localized modes remain with large eigenvectors in the local regions. In fact, in larger simulation cells, the majority of the highly contributing modes to the local ion hop become the localized modes, because of their large eigenvectors of vibration. Therefore, by virtue of exciting localized modes, the power density reduces as we increase the system size. However, additional simulations at larger system sizes have shown that the contribution of excited localized modes to the atom hopping remains virtually unchanged even in larger simulation cells, as is evidenced by calculating the ion jump (hop) rate. As Table S5 shows if we excite the highest singly contributing mode to the atom hop in the system, the jump rate stays almost constant as we increase the system size, yet the power density decreases with increasing size. Therefore, a high power density need not be necessary to realize this. Table S5 also shows how low this power density would be if such a small amount of power were extrapolated to a macroscopic sized system. The power density, which in essence would manifest itself like heat generation is only on the order of 10 W/m$^3$. To put this in perspective, consider the heat generation rate that occurs in 12 guage copper wires commonly used in residential wiring, which typically sees currents ranging from 1-10 Amps (limit ~

40A). For such wires, which do not heat up appreciably, and are able to easily cool themselves via natural convection and radiation to the surroundings, currents ranging from 1-10 Amps (e.g., a 100 W light bulb vs. a toaster), the corresponding heat generation rates would be ~1.58-158 kW/m3. Thus, one could afford to increase the power density of targeted mode excitation by 2-4 orders of magnitude from the extrapolated estimation based on the power inputs to the small supercells studied herein.

**Table S5.** By increasing the system size, jump rate stays almost, but the required power to drive the hop decreases.

| System size | 1x (192 atoms) | 2x (384 atoms) | 3x (576 atoms) | Bulk (1g) $3 \times 10^{22}$ atoms $Li_3PO_4$ |
|---|---|---|---|---|
| Jump rate (#jumps/ps) | 0.241 | 0.237 | 0.234 | |
| Required power (W/m$^3$) | $1.2 \times 10^{19}$ | $6.7 \times 10^{18}$ | $3.8 \times 10^{18}$ | $1.22 \times 10^{1}$ |
| Required power (W/g) | | | | $2.4 \times 10^{-3}$ |
| Heat capacity of $Li_3PO_4$ (J/g-K) | | | | 1.25 |

The modal excitations in the performed simulations in this manuscript were done every 5 time steps (every 5 fs). If the excitations rate is decreased from every 20 fs to lower values, such as 100 fs, the increase in hopping rate and diffusivity will diminish. In particular, every 5 fs excitation, cannot be sustained with the existing THz laser technologies with their typical short duty cycles and a cycle repetition period of around milliseconds.[120] However, continuous excitations can still be achieved using continuous wave lasers and indirectly coupling their photons with the highest contributing modes in the system, and thus sustaining the excitations.